\documentclass[showpacs,preprintnumbers]{revtex4}%
\usepackage{hyperref}
\usepackage{amsfonts}
\usepackage{amsmath}
\usepackage{amssymb}
\usepackage{graphicx}%

\begin{document}
\title{Energy and entropy radiated by a black hole embedded in the de-Sitter braneworld}
\author{Shao-Feng Wu$^{1,2}$\footnote{Corresponding author. Email: sfwu@shu.edu.cn;
Phone: +86-021-66136202.}, Shaoyu Yin$^{3}$, Guo-Hong Yang$^{1,2}$, and
Peng-Ming Zhang$^{4,5}$}
\affiliation{$^{1}$Department of Physics, College of Science, Shanghai University,
Shanghai, 200444, P. R. China}
\affiliation{$^{2}$The Shanghai Key Lab of Astrophysics, Shanghai, P. R. China}
\affiliation{$^{3}$Department of Physics, Fudan University, Shanghai 200433, P. R. China}
\affiliation{$^{4}$Center of Theoretical Nuclear Physics, National Laboratory of Heavy Ion
Accelerator, Lanzhou 730000, P. R. China}
\affiliation{$^{5}$Institute of Modern Physics, Lanzhou, 730000, P. R. China}

\begin{abstract}
We study the Hawking radiation of (4+n)-dimensional Schwarzschild black hole
imbedded in the space-time with positive cosmological constant. The greybody
and energy emission rates of scalars, fermions, bosons, and gravitons are
calculated in the full range of energy. The valuable information on the
dimensions and curvature of space-time is revealed. Furthermore, we
investigate the entropy radiated and lost by black hole. We find their ratio
near unit in favor of the Bekenstein's conjecture.

\end{abstract}
\maketitle

\section{Introduction}

The concept of extra dimensions is becoming increasingly popular in both
particle theory, such as the string theory and the grand unification, and
cosmology. It also suggests us a new solution to the hierarchy problem, since
the traditional Planck scale can be derived as an effective energy scale from
the fundamental higher-dimensional scale \cite{ADD,Randall}. This idea opens
possibilities for observing strong gravitational phenomena if the fundamental
Planck scale is lowered to the order of magnitude around TeV. Based on the
well-known braneworld scenario with large or compact extra dimensions, a
particularly exciting proposal is the possibility of mini black holes with
horizon radii smaller than the size of the extra dimensions centered on our
braneworld and extending along the extra dimensions. Even the future
possibility has been discussed to produce such mini black holes in super
collider, with the center-of-mass energy greater than the fundamental scale
\cite{Banks}. It has also been expected to observe the mini black holes via
the ultrahigh-energy neutrinos interacting with the atmosphere of the earth
\cite{Goyal}. Moreover, the mini black holes might have been created in the
early universe due to density perturbations or phase transitions. The
observational signatures of such mini black holes can be precisely determined
by their well-known quantum gravity effect, the Hawking radiation, which
encodes the vital information about the structure of the space-time
geometrical background. Much effort on the Hawking radiation has been expanded
to study the properties of the extra dimensions of space-time \cite{Kanti},
the string-induced Gauss-Bonnet correction \cite{Barrau,Grain}, and the
cosmological constant \cite{SdS}. The attention on the de Sitter space-time is
motivated at least by the following three aspects. First, it is important to
study non-asymptotically flat space-time since there is a duality between
quantum gravity on the (A)dS space and the Euclidean conformal field theory
(CFT) \cite{Maldacen}. Second, recent astronomical observations on supernova
indicate that the present universe be dominated by energy with negative
pressure \cite{Riess07}, and a positive cosmological constant is a ready
candidate. Third, it is generally accepted that the evolution of the universe
is well represented by the FRW cosmology punctuated by a de Sitter-like
exponentially inflationary phase \cite{Guth}. Therefore, it is reasonable to
ask what effect the inflation might have on the behavior of the primordial
black hole.

In this paper, we will study the Hawking radiation of the black hole imbedded
in higher-dimensional space-time with the asymptotical dS boundary. Compared
with the vast literatures on Hawking radiation, the work accounting for
Schwarzschild-de-Sitter (SSdS) black hole is much sparser. For the
four-dimensional case, after the pioneer work of Gibbons and Hawking
\cite{Gibbons}, the existing literatures mainly considered the two-dimensional
Vaidya-de Sitter space-time \cite{Mallett}, or focused on the evaporating
dilatonic final state of a SSdS black hole \cite{Bousso}. For the
higher-dimensional case, most activities have been focused on the study of the
pair creation of black holes \cite{Dias}, thermodynamical radiation via
tunneling \cite{Medved}, the mass and entropy bounds \cite{Boer}, and the
quasinormal frequencies associated with the perturbations of the
higher-dimensional space-time \cite{Molina,Natario,Konoplya}. Some possible
experimental consequences of the higher-dimensional dS or AdS evaporating
black holes have been preliminary studied in Ref. \cite{Labbe}. The exact form
of the higher-dimensional Hawking radiation spectrum was first studied in Ref.
\cite{SdS}, where only scalar fields are considered. But it is still unknown
for fermions, gauge bosons, and gravitons including tensor, vector, and scalar
modes. The gravitons recently have been analytically considered but only at
low or at large imaginary frequency \cite{Harmark}. One aim of our work is to
give the exact energy spectrum radiated by higher-dimensional SSdS black hole
and examine the effect of the dimensions of space-time and the cosmological constant.

Since the key ingredient for the energy spectrum is the greybody factor, or
equally, the absorption probability, which distinguishes the black hole
radiation from the black body, we should know the greybody factor of
higher-dimensional SSdS black hole before we get the energy spectrum. With the
spectrum of greybody factor at hand, we will investigate the entropy radiation
of higher-dimensional SSdS black hole. It is well-known that one of the most
enchanting features of black hole thermodynamics is the Bekenstein's
conjecture \cite{Bekenstein}, positing that a black hole possesses an entropy
proportional to its surface area, which can be statistically interpreted as a
measure of all possible pre-collapse configurations before the black hole was
built. Recovering this entropy is an important success of the quantum gravity
\cite{Ashtekar} and closely connects to the proposal of the (A)dS/CFT
correspondence. However, a direct proof of the statistical interpretation of
the entropy is still missing. Under the consideration that the evaporation of
a black hole is a time-reversal process of its collapse, and that the entropy
of the final state of black hole can be accurately estimated, Zurek
\cite{Zurek} proposed that Bekenstein's conjecture can be tested by
quantifying the ratio of the evaporated entropy to the entropy lost by the
black hole. By numerical calculations based on the greybody factor of
four-dimensional Schwarzschild black hole, Page \cite{Page} showed the ratio
is indeed near unit, which is strongly in favor of the conjecture of
Bekenstein, since nothing mathematically prevents the ratio being arbitrarily
large or small\textbf{.} The argument of Zurek was immediately generalized to
the four-dimensional Schwarzschild black hole with charge \cite{Schumacher}.
However, because of the sophisticated numerical techniques needed in the
greybody factor calculation, similar study on other black hole, especially the
higher-dimensional black hole, is not present until very recent work of Barrau
et al. \cite{Barrau1}. We now wish to extend Zurek's argument to the case of
higher-dimensional SSdS black hole. Quite different from the case of the flat
space-time, the greybody factor for scalar field has already been found
non-vanishing in the low energy limit \cite{SdS}, and we obtain the similar
result for fermion fields. Besides, we also study the behavior of vector and
tensor fields for completeness.

The organization of the paper is as follows: In section II, we briefly review
the general framework of SSdS black hole. In section III, we give the
analytical solutions of master equations for all particles near\ the event
horizon and the cosmological horizon. In section IV, we concentrate on the
Hawking radiation of a decaying SSdS black hole. We will show the exact
greybody, energy spectrum and entropy variation of all particle species, with
the strong effect of the extra dimensions and the cosmological constant. The
last section is devoted to conclusions and discussions.

\section{SSdS Black-Hole Properties}

\begin{figure}[ptb]
\begin{center}
\includegraphics[
height=2in, width=4in]{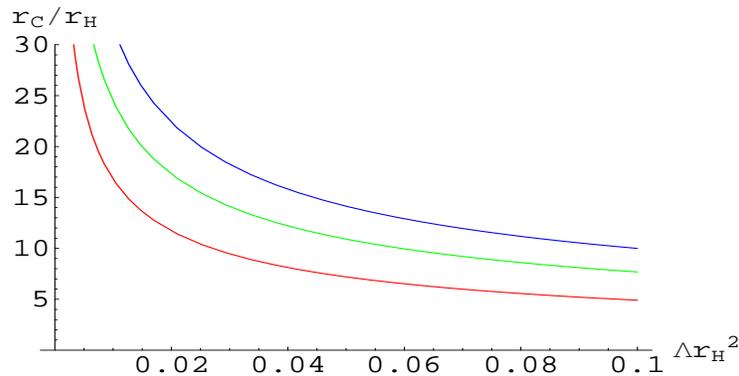}
\end{center}
\caption{The ratio between the cosmological horizon and event horizon
$r_{C}/r_{H}$ with respect to the (dimensionless) cosmological constant
$\Lambda r_{H}^{2}$ for $n=0\;$(red),\ 1\ (green),\ 2 (blue).}%
\label{fig1}%
\end{figure}

We consider the class of black holes that are formed in the presence of a
positive cosmological constant $\Lambda$ in the $d$-dimensional $(d=4+n)$
space-time. The geometrical background of SSdS black holes is given by the
Tangherlini line element \cite{Tangherlini}%
\begin{equation}
ds^{2}=-f(r)dt^{2}+\frac{dr^{2}}{f(r)}+r^{2}d\Omega_{n+2}^{2}
\label{high line}%
\end{equation}
where%
\begin{equation}
f(r)=1-\frac{\mu}{r^{n+1}}-\frac{2\kappa_{d}^{2}\Lambda r^{2}}{(n+2)(n+3)}.
\label{hr}%
\end{equation}
The $d\Omega_{n+2}$ is the solid angle element. The parameter $\mu$ is related
to the ADM mass of the black hole through the relation%
\[
M=\frac{(n+2)A_{n+2}\mu}{2\kappa_{d}^{2}},
\]
where $A_{n+2}$ is the area of a unit ($n+2$)-dimensional sphere. The
$\kappa_{d}^{2}=8\pi G_{d}$ stands for the $d$-dimensional Newton's constant,
which will be set to unit for convenience. There are two positive roots of
equation $f(r)=0$, the larger one ($r_{C}$) corresponding to the cosmological
horizon, and the smaller one ($r_{H}$) to the event horizon. From Fig.
\ref{fig1}, one can find that the cosmological horizon will be close to the
event horizon when the cosmological constant is big and the space-time
dimensions are small. When the two horizons are lying close to each other,
Nariai black hole will be arisen, which corresponds to the maximum black hole
and minimum de Sitter space \cite{Nariai}. 

Using the relation between the parameter $\mu$ and two horizons%
\begin{equation}
\mu=r_{H/C}^{n+1}\left[  1-\frac{2\Lambda}{(n+2)(n+3)}r_{H/C}^{2}\right]  ,
\label{u}%
\end{equation}
we can relate the metric with two horizons%
\begin{equation}
f(r)=1-\frac{2\Lambda r^{2}}{(n+2)(n+3)}+\frac{r_{H/C}^{n+1}}{r^{n+1}}\left[
\frac{2\Lambda}{(n+2)(n+3)}r_{H/C}^{2}-1\right]  , \label{fr}%
\end{equation}
and the ADM mass with the event horizon%
\begin{equation}
M=\frac{(n+2)A_{n+2}}{2}r_{H}^{n+1}\left[  1-\frac{2\Lambda}{(n+2)(n+3)}%
r_{H}^{2}\right]  . \label{MH}%
\end{equation}

\begin{figure}[ptb]
\begin{center}
\includegraphics[
height=2in, width=6in]{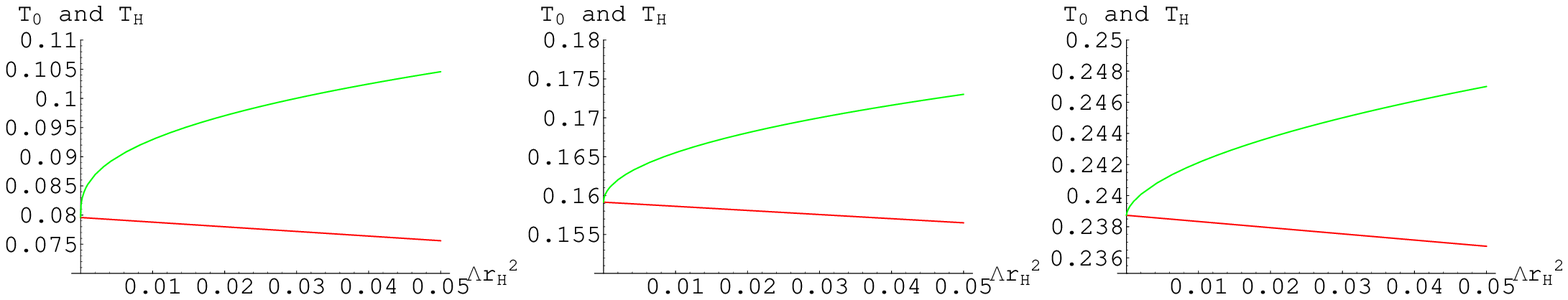}
\end{center}
\caption{Temperature $T_{0}$ (red) and $T_{H}$ (green), in units of
$r_{H}^{-1}$, with respect to the cosmological constant $\Lambda r_{H}^{2}$,
for $n=0\;$(left),\ 1\ (center),\ 2 (right), notice that the scale of these
three plots are different.}%
\label{fig2}%
\end{figure}

The temperature of SSdS black hole, defined by the surface gravity, is rather
subtle. There are two kinds of temperatures based on the standard and
Bousso-Hawking normalizations, respectively. The standard normalization
provides the Hawking temperature $T_{0}$, which was derived by the analogy
with asymptotically flat space-time \cite{Labbe,Harmark,Myung1}. In the case
of asymptotically flat space-time, the standard method to obtain the surface
gravity is to choose the normalized Killing field at infinity, where an
observer does not feel any acceleration. However, in the presence of a
cosmological constant, there is no asymptotic flat region and we can not place
the preferred observer at infinity. Usually, the Bousso-Hawking normalization
is introduced \cite{Bousso1,SdS}. Consider the zero point $r_{0}$ of the first
derivative of metric function $f^{\prime}(r)$,
\begin{equation}
r_{0}=\left[  \frac{(n+1)(n+2)(n+3)\mu}{4\Lambda}\right]  ^{\frac{1}{n+3}},
\label{r0}%
\end{equation}
at which the metric function achieves the maximum value. The black hole
attraction and the cosmological repulsion exactly cancel out at this point,
thus one may achieve the zero acceleration inside the cosmological horizon.
Now we can obtain the Bousso-Hawking temperature of the black hole%
\[
T_{H}=\frac{1}{\sqrt{f(r_{0})}}T_{0},
\]
where%
\begin{equation}
T_{0}=\frac{1}{4\pi r_{H}}\left[  (n+1)-\frac{2\Lambda}{n+2}r_{H}^{2}\right]
\label{T0}%
\end{equation}
is the Hawking temperature obtained by the standard normalization.
Substituting Eq. (\ref{u}) and Eq. (\ref{r0}) into\ Eq. (\ref{hr}), we can
plot the two temperatures $T_{H}$ and $T_{0}$ as the function of extra
dimensions $n$ and the cosmological constant $\Lambda r_{H}^{2}$, see Fig.
\ref{fig2}. One can find that the standard Hawking temperature $T_{0}$
increases with $n$ but decreases with $\Lambda r_{H}^{2}$; however, both
parameters cause an increase in the Bousso-Hawking temperature $T_{H}$. In
most part of this paper, we apply the Bousso-Hawking temperature following the
work of Ref. \cite{SdS}. However, we want to emphasize that the greybody
factor and the method to obtain the energy and entropy radiation are not
dependent on the different choice of temperatures. 

The entropy of the black hole is proportional to its surface area%
\[
S=\frac{1}{4G}A_{n+2}r_{H}^{n+2}.
\]
Using Eqs. (\ref{MH}) and (\ref{T0}), we can obtain the loss of entropy%
\[
dS=\frac{1}{T_{0}}dM.
\]

With a non-vanishing temperature, SSdS black hole will emit Hawking radiation
in the form of elementary particles, which is similar as the blackbody
radiation, with a density matrix in each mode \cite{Wald}%
\[
\rho_{kk^{\prime}}=\delta_{kk^{\prime}}\frac{\left\vert A_{j}^{(s)}\right\vert
^{2k}\left[  e^{\frac{\omega}{T_{H}}}-\left(  -1\right)  ^{2s}\right]
}{\left[  e^{\frac{\omega}{T_{H}}}-\left(  -1\right)  ^{2s}+\left(  -1\right)
^{2s}\left\vert A_{j}^{(s)}\right\vert ^{2k}\right]  ^{k+\left(  -1\right)
^{2s}}},
\]
where $s$ denotes the spin of particle, $j$ is the angular momentum number,
and $\left\vert A_{j}^{(s)}\right\vert ^{2}$ is the greybody factor. The
energy emission rate is $Tr(-\rho\omega)$, which gives%
\begin{equation}
dE^{(s)}=\frac{1}{2\pi}\sum_{j}N_{j}^{(s)}\left\vert A_{j}^{(s)}\right\vert
^{2}\frac{\omega}{e^{\frac{\omega}{T_{H}}}-\left(  -1\right)  ^{2s}}d\omega,
\label{dEs}%
\end{equation}
where $N_{j}^{(s)}$ are the multiplicities of states that correspond to the
angular momentum number $j$. For fields localized on brane, the multiplicities
of states are $N_{j}^{(s)}=2j+1$, while for gravitons, the multiplicities of
states are different for three types of perturbation \cite{Rubin}
\[
N_{j}^{(s)}=\frac{(2j+d-3)(j+d-4)}{j!(d-3)!}\;\text{for scalar mode,}%
\]%
\[
N_{j}^{(s)}=\frac{j(j+d-3)(2j+d-3)(j+d-5)!}{(j+1)!(d-4)!}\;\text{for vector
mode,}%
\]%
\[
N_{j}^{(s)}=\frac{(d-4)(d-1)(j+d-2)(j-1)(2j+d-3)(j+d-5)!}{2(j+1)!(d-3)!}%
\;\text{for tensor mode.}%
\]
Integrating this energy spectrum over all modes gives the exact expression of
entropy loss of a black hole%
\begin{equation}
dS=\frac{1}{T_{0}}dM=\sum_{s,p,j}N_{j}^{(s)}\int_{0}^{\infty}d\omega
h_{j}^{(s)}(w), \label{dS}%
\end{equation}
where $p$ denotes the helicity, and the integrand of the lost entropy
$h_{j}^{(s)}(w)$ is%
\begin{equation}
h_{j}^{(s)}(w)=\frac{1}{T_{0}}\frac{1}{2\pi}\left\vert A_{j}^{(s)}\right\vert
^{2}\frac{\omega}{e^{\frac{\omega}{T_{H}}}-\left(  -1\right)  ^{2s}}.
\label{h}%
\end{equation}
On the other hand, the entropy emission rate $Tr(-\rho\ln\rho)$ is expressed
as
\begin{equation}
dS_{rad}=\sum_{s,p,j}N_{j}^{(s)}\int_{0}^{\infty}d\omega g_{j}^{(s)}(w),
\label{dSrad}%
\end{equation}
where the integrand of radiated entropy $g_{j}^{(s)}(w)$ is
\begin{equation}
g_{j}^{(s)}(w)=\frac{1}{2\pi}\left\{  \frac{\left\vert A_{j}^{(s)}\right\vert
^{2}}{e^{\frac{\omega}{T_{H}}}-\left(  -1\right)  ^{2s}}\ln\left[
\frac{e^{\frac{\omega}{T_{H}}}-\left(  -1\right)  ^{2s}}{\left\vert
A_{j}^{(s)}\right\vert ^{2}}+\left(  -1\right)  ^{2s}\right]  +\left(
-1\right)  ^{2s}\ln\left[  1+\left(  -1\right)  ^{2s}\frac{\left\vert
A_{j}^{(s)}\right\vert ^{2}}{e^{\frac{\omega}{T_{H}}}-\left(  -1\right)
^{2s}}\right]  \right\}  . \label{g}%
\end{equation}
For our aim, we will evaluate the energy spectrum (\ref{dEs}) and the ratio of
the entropy lost by the black hole to the entropy gained by the radiation%
\[
R=\frac{dS_{rad}}{dS}.
\]
Obviously, it is necessary to calculate the greybody $\left\vert A_{j}%
^{(s)}\right\vert ^{2}$ at first. It is computed within a full quantum
mechanical framework in the aforementioned SSdS black hole background.

\section{Master equations and their solutions near horizons}

In this section, we will give the master equations for all elementary
particles and the solutions near the event horizon and the cosmological horizon.

\subsection{brane particles}

In the braneworld scenarios, standard model fields are confined on the brane,
except the gravitons which can propagate in the extra dimensions. We first
consider the master equation for particles on the brane. Particles confined on
the brane propagated in a background whose geometry is induced by the bulk
curvature. The induced geometry on the 4-dimensional brane is given by fixing
the values of the extra azimuthal angular coordinates and leads to the
projection of the $d$-dimensional metric on the 4-dimensional slice that
describes our world%
\[
ds^{2}=-f(r)dt^{2}+\frac{dr^{2}}{f(r)}+r^{2}d\Omega_{2}^{2}.
\]
Here metric function remains the same as in the higher-dimensional line
element (\ref{hr}), so the profile of the curvature along the three noncompact
spatial dimensions contains a fingerprint of the bulk curvature. Since the
line element of a $d$-dimensional SSdS black hole has a Schwarzschild-like
form, the motion equations for brane particles can be described by the well
known Newman-Penrose formalism \cite{Newman}. Factorizing the propagating
field as%
\[
\Psi_{s}(t,r,\theta,\varphi)=e^{-i\omega t}\Delta^{-s}P_{s}(r)Y_{s,j}%
(\theta,\varphi),
\]
where $\Delta=fr^{2}$, $Y_{s,j}$ is the spin-weighted spherical harmonics, the
radial part of the master equation of motion has the form \cite{Ida,Goldberg}%
\begin{equation}
\Delta^{s}\frac{d}{dr}(\Delta^{1-s}\frac{dP_{s}}{dr})+(\frac{\omega^{2}r^{2}%
}{f}+2is\omega r-\frac{is\omega r^{2}f^{\prime}}{f}-\bar{\lambda})P_{s}=0,
\label{NP}%
\end{equation}
where $\bar{\lambda}=j(j+1)-s(s-1)$ and the prime denotes the derivative with
respect to $r$. The master equation can be rewritten as the familiar
Shr\"{o}dinger-like form%
\[
-\frac{d^{2}U}{dr_{\ast}^{2}}+VU=\omega^{2}U,
\]
where $r_{\ast}$ is the tortoise coordinate defined by $dr_{\ast}=dr/f$,
$U=r^{1-s}f^{-s/2}P_{s}$ is the redefined radial function, and the potential
is%
\begin{equation}
V=\frac{is\omega\left(  -2f+rf^{\prime}\right)  }{r}+\frac{4s(s-1)f^{2}%
+r^{2}s^{2}f^{\prime2}+f(4\bar{\lambda}+4r(s-1)^{2}f^{\prime}-2sr^{2}%
f^{\prime\prime})}{4r^{2}}. \label{V}%
\end{equation}
To find the absorption coefficient for propagation of fields on the brane, we
need to solve the radial equation (\ref{NP}) over the whole radial regime.
However, even in the absence of a bulk cosmological constant, the general
solution of this equation is extremely difficult to be found. It may be
analytically solved with some approximations (at the low or high-energy
regime). However, a complete solution for general gravitational background
inevitably requires numerical methods. To solve numerically the master
equation and extract the absorption coefficient under the Schwarzschild
background, one needs to check the analytical behavior near the event horizon
and infinity; while for the SSdS case, the solution near the cosmological
horizon should be considered instead. A simple method will be given to obtain
the desired asymptotic solutions. The key point of this method is to express
the master equation using the two horizons, respectively. Each equation is
used to solve the corresponding asymptotic solution. Substituting the metric
functions expressed by two horizons (\ref{fr}) into the master equation
(\ref{NP}) respectively, we have%
\[
C_{2}(r_{H/C})P_{s}^{\prime\prime}+C_{1}(r_{H/C})P_{s}^{\prime}+C_{0}%
(r_{H/C})P_{s}=0,
\]
where $C_{i}(r_{H/C})$ are the functions of event horizon/cosmological
horizon. In order to find the suitable asymptotic equations which can be
analytically solved, we expand the functions $C_{i}(r_{H/C})$ to the lowest
non-vanishing order of $r-r_{H/C}$:
\[
C_{2}^{0}(r_{H/C})=-\left[  3-r_{H/C}^{2}K^{2}+d(r_{H/C}^{2}K^{2}-1)\right]
^{2}\left(  r-r_{H/C}\right)  ^{2}%
\]%
\[
C_{1}^{0}(r_{H/C})=(s-1)\left[  3-r_{H/C}^{2}K^{2}+d(r_{H/C}^{2}%
K^{2}-1)\right]  ^{2}\left(  r-r_{H/C}\right)
\]%
\[
C_{0}^{0}(r_{H/C})=-ir_{H}\omega\{s\left[  3-r_{H/C}^{2}K^{2}+d(r_{H/C}%
^{2}K^{2}-1)\right]  -ir_{H/C}\omega\},
\]
where $K^{2}=\frac{2\Lambda}{(n+2)(N+3)}$. Then the asymptotic equations are%
\begin{equation}
C_{2}^{0}(r_{H/C})P^{\prime\prime}+C_{1}^{0}(r_{H/C})P^{\prime}+C_{2}%
^{0}(r_{H/C})P=0, \label{as eq}%
\end{equation}
with solutions%
\begin{equation}
P_{s}(r)=A_{1}(r-r_{H})^{\frac{-i\omega r_{H}}{d-3-(d-1)r_{H}^{2}K^{2}}%
},\;\text{for }r\gtrsim r_{H} \label{xh}%
\end{equation}%
\begin{equation}
P_{s}(r)=B_{1}(r-r_{C})^{\frac{-i\omega r_{C}}{d-3-(d-1)r_{C}^{2}K^{2}}}%
+B_{2}(r-r_{C})^{\frac{i\omega r_{C}}{d-3-(d-1)r_{C}^{2}K^{2}}+s},\;\text{for
}r\lesssim r_{C}. \label{cx}%
\end{equation}
In Eq. (\ref{xh}) we keep only the incoming modes in order to satisfy the
boundary condition at the black hole event horizon, while at the cosmological
horizon in Eq. (\ref{cx}), both incoming and outgoing modes can exist. We will
see that Eqs. (\ref{xh}) and (\ref{cx}) are suitable to the numerical
computation for all fields on the brane.

\subsection{gravitons}

A graviton can propagate in the bulk, so the gravitational background is
described by the complete line element expressed in Eq. (\ref{high line}).
Decomposing the graviton into a symmetric traceless tensor, a vector and a
scalar part, Kodama and Ishibashi \cite{Ishibashi} have obtained the master
equations for the Schwarzschild-like line element
\begin{equation}
-\frac{d^{2}\Phi}{dr_{\ast}^{2}}+V\Phi=\omega^{2}\Phi. \label{KI}%
\end{equation}
The potential $V$ has a different form for each type of perturbation, namely%
\[
V_{T}=f(r)\left[  \frac{(d-2)(d-4)f(r)}{4r^{2}}+\frac{j(j+d-3)}{r^{2}}%
+\frac{(d-2)f^{\prime}(r)}{2r}\right]
\]
for tensor perturbation,
\[
V_{V}=f(r)\left[  \frac{(d-2)(d-4)f(r)}{4r^{2}}+\frac{j(j+d-3)}{r^{2}}%
-\frac{(d-2)f^{\prime\prime\prime}(r)}{2r(d-3)}\right]
\]
for vector perturbation, and%
\[
V_{S}=\frac{f(r)}{r^{2}}\frac{\left(  q\alpha^{3}+p\alpha^{2}+w\alpha
+z\right)  }{\left[  4m+2(d-1)(d-2)\alpha\right]  ^{2}}%
\]
for scalar perturbation where%
\[
m=j(j+d-3)-(d-2)
\]%
\[
\alpha=\frac{\mu}{r^{d-3}}%
\]%
\[
q=(d-2)^{4}(d-1)^{2}%
\]%
\[
p=(d-2)(d-1)\left[  (d-6)(d-4)(d-2)(d-1)+4(2d^{2}-11d+18)m\right]
-2(d-2)^{2}(d-1)d\Lambda r^{2}%
\]%
\[
w=-12(d-2)m\left[  (d-4)(d-2)(d-1)+(d-6)m\right]  +24(d-4)(d-2)m\Lambda r^{2}%
\]%
\[
z=4(d-2)dm^{2}+16m^{3}-\frac{8(d-6)(d-4)m^{2}\Lambda r^{2}}{(d-2)(d-1)}.
\]
One can find that three potentials have the same form as the ones for
Schwarzschild black hole \cite{Creek,Cardoso} except the additional terms with
$\Lambda$ in $p$, $w$, and $z$. The method to obtain the asymptotic equations
is similar to the aforementioned one. Interestingly, we find that three
different potentials have the same asymptotic behaviour. The asymptotic
equations are similar as Eq. (\ref{as eq}) but with different functions
$C_{i}^{0}(r_{H/C})$%
\[
C_{2}^{0}(r_{H/C})=\left[  3-r_{H/C}^{2}K^{2}+d(r_{H/C}^{2}K^{2}-1)\right]
^{2}(r-r_{H/C})^{2}%
\]%
\[
C_{1}^{0}(r_{H/C})=\left[  3-r_{H/C}^{2}K^{2}+d(r_{H/C}^{2}K^{2}-1)\right]
(r-r_{H/C})
\]%
\[
C_{0}^{0}(r_{H/C})=\omega^{2}r_{H/C}^{2}.
\]
The corresponding solutions are%
\begin{equation}
\Phi(r)=E_{1}(r-r_{H})^{-i\frac{\omega r_{H}}{d-3-(d-1)r_{H}^{2}K^{2}}%
},\;\text{for }r\gtrsim r_{H} \label{xh1}%
\end{equation}%
\begin{equation}
\Phi(r)=F_{1}(r_{C}-r)^{-i\frac{\omega r_{C}}{d-3-(d-1)r_{C}^{2}K^{2}}}%
+F_{2}(r_{C}-r)^{i\frac{\omega r_{C}}{d-3-(d-1)r_{C}^{2}K^{2}}},\;\text{for
}r\lesssim r_{C}. \label{cx1}%
\end{equation}

\section{Energy spectrum and entropy variation}

To evaluate the absorption spectrum in a wide energy range accurately, it is
necessary to turn to numerical calculations. The numerical integration of
master equations (\ref{NP}) and (\ref{KI}) is performed from the black-hole
event horizon, where appropriate boundary conditions (\ref{xh}) and
(\ref{xh1}) are applied, and extends to the cosmological horizon. Then the
absorption coefficient can be extracted by fitting the analytical asymptotic
solutions (\ref{cx}) and (\ref{cx1}) to the numerical results. In the
following, we will evaluate the absorption probability, energy spectrum and
entropy variation of each particle species.

\subsection{Scalar field ($s=0$)}

\begin{figure}[ptb]
\begin{center}
\includegraphics[
height=3in, width=6in ]{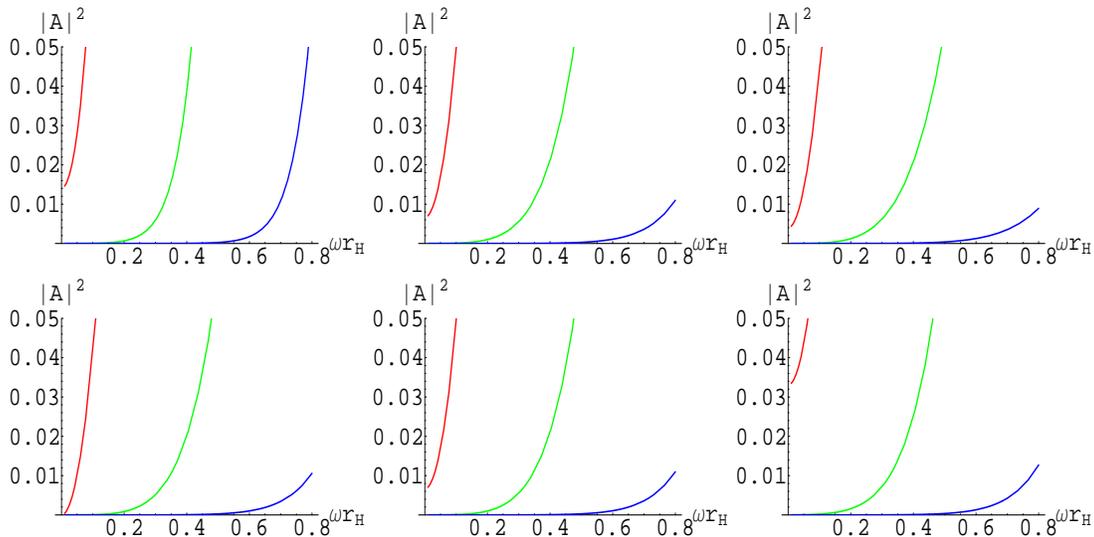}
\end{center}
\caption{Absorption probability $\left\vert A_{j}^{(0)}\right\vert ^{2}$ of
scalar field versus the parameter $\omega r_{H}$, with $j$ $=0$ (red)$,\;1$
(green)$,\;2$\ (blue). For the top panel: $\Lambda r_{H}^{2}=0.01,$ and
$n=0,\;1,\;2$ from left to right; for the bottom panel: $n=1$ and $\Lambda
r_{H}^{2}=0,\;0.01,\;0.05$ from left to right.}%
\label{fig3}%
\end{figure}

\begin{figure}[ptb]
\begin{center}
\includegraphics[
height=2in, width=6in ]{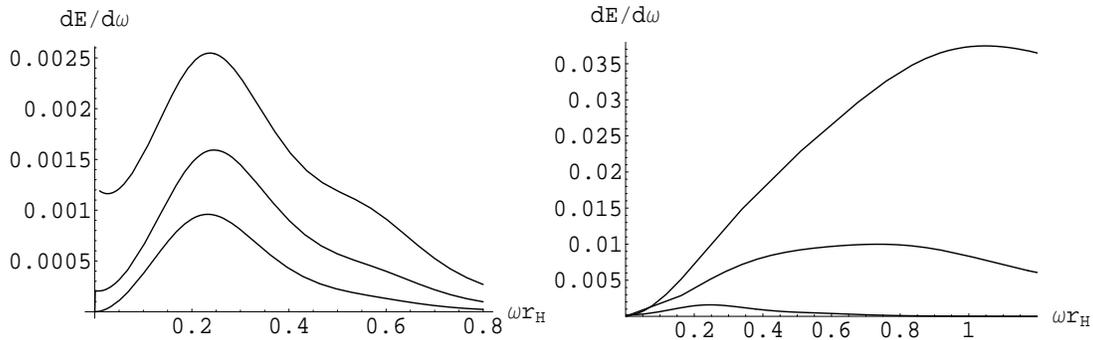}
\end{center}
\caption{The differential energy emission rate, for scalar emission on the
brane, versus the parameter $\omega r_{H}$. For the left panel: the
dimensionality of space-time is fixed at $n=0$, while $\Lambda r_{H}^{2}$
takes the values $\{0,0.01,0.05\}$ from bottom to top; for the right panel:
the cosmological constant is fixed at $\Lambda r_{H}^{2}$ $=0.01$, and $n$
takes the values $\{0,2,4\}$ from bottom to top.}%
\label{fig4}%
\end{figure}

The absorption probability is defined as the ratio of the ingoing flux at two
horizons. For scalar particles, the flux is \cite{Grain}%
\[
F=2fr^{2}\operatorname{Im}\left[  \frac{P_{0}}{r^{2}}\frac{d}{dr}\left(
r^{2}P_{0}^{+}\right)  \right]  .
\]
The absorption probability is found as%
\[
\left\vert A_{j}^{(0)}\right\vert ^{2}=1-\left\vert \frac{B_{2}}{B_{1}%
}\right\vert ^{2}.
\]
Now we introduce the numerical method to obtain $\left\vert A_{j}%
^{(0)}\right\vert ^{2}$ following Ref. \cite{Jung}. The asymptotic solution
near the cosmological horizon (\ref{cx}) can be written as%
\[
P_{0}(x)=B_{1}F_{-}+B_{2}F_{+},
\]
where $F_{-}$ and $F_{+}$ denote the ingoing and outgoing waves, respectively.
Consider the Wronskians of $F^{-}$ and $F^{+}$, which have the following
property%
\[
W\left[  F_{-},P_{0}\right]  \equiv F_{-}P_{0}^{\prime}-P_{0}F_{-}^{\prime
}=B_{2}W\left[  F_{-},F_{+}\right]
\]%
\[
W\left[  F_{+},P_{0}\right]  \equiv F_{+}P_{0}^{\prime}-P_{0}F_{+}^{\prime
}=-B_{1}W\left[  F_{-},F_{+}\right]  .
\]
By solving the radial equation numerically and comparing two Wronskians, we
obtain the absorption probability%
\[
\left\vert A_{j}^{(0)}\right\vert ^{2}=1-\left\vert \frac{W\left[  F_{-}%
,P_{0}\right]  }{W\left[  F_{+},P_{0}\right]  }\right\vert ^{2}.
\]
In Fig. \ref{fig3}, we depict the absorption probability of scalar field with respect
to the energy parameter $\omega r_{H}$, for different angular momentum $j$,
extra dimensions $n$, and the parameter of cosmological constant $\Lambda
r_{H}^{2} $. It is important to note that for the lowest mode $\left(
j=0\right)  $\ the absorption probability is nonvanishing in the presence of
cosmological constant. This is definitely different from the case of
Schwarzschild black hole. As pointed out in Ref. \cite{SdS}, the finite
absorption coefficient in the infrared limit can be understood in the
following way: in the presence of a second horizon, our universe is within the
two horizon and therefore has only finite size, in which the particle
propagates with infinite wavelength cannot be localized, and thus has a finite
probability to propagate through the barrier and be absorbed in the black
hole. The energy spectrum is given in Fig. \ref{fig4}, which is the same as
that obtained by Ref. \cite{SdS}. 

\begin{figure}[ptb]
\begin{center}
\includegraphics[
height=3in, width=6in ]{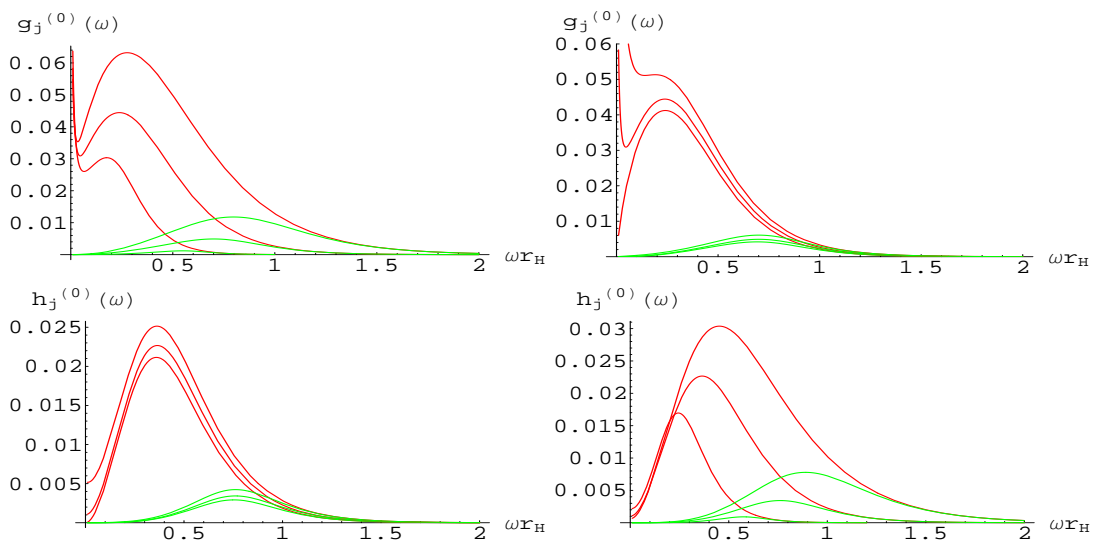}
\end{center}
\caption{The integrands of entropy variation, for scalar emission on the
brane, versus the parameter $\omega r_{H}$, with $j$ $=0$ (red) and$\;1$
(green). The top left panel shows the integrand of radiated entropy
$g_{j}^{(0)}(w)$, with $\Lambda r_{H}^{2}=0.01,$ and $n=0,\;1,\;2$ from bottom
to top; while in the top right panel, $n=1$, and $\Lambda r_{H}^{2}%
=0,\;0.01,\;0.05$ from bottom to top. The corresponding integrand of lost
entropy $h_{j}^{(0)}(w)$ are given in the bottom panel. One can find that the
lowest mode of integrand $g_{0}^{(0)}(w)$ diverges as $\omega r_{H}%
\rightarrow0$. But for higher modes or the lost entropy, \ the integrand is
finite.}%
\label{fig5}%
\end{figure}

\begin{figure}[ptb]
\begin{center}
\includegraphics[
height=2in, width=6in ]{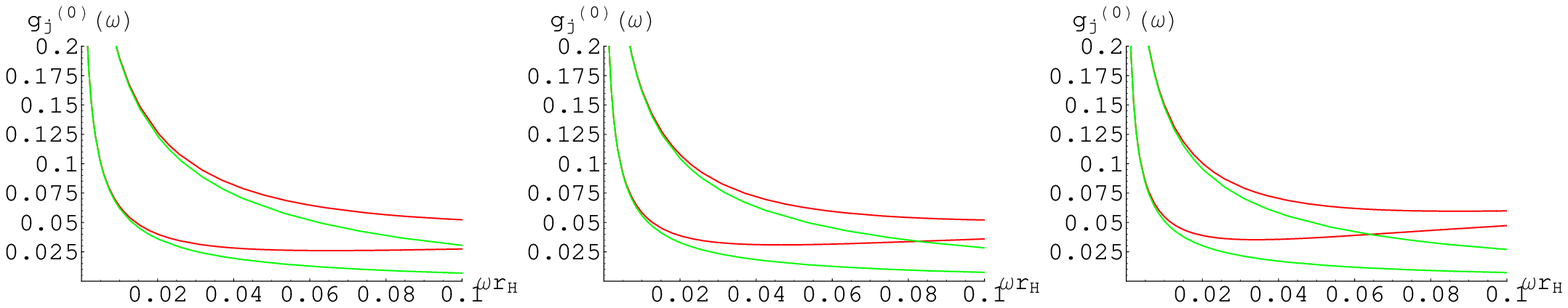}
\end{center}
\caption{The numerical (red) and analytical (green) zero mode of integrated
function of radiated entropy of scalar field $g_{0}^{(0)}(w)$, with $\Lambda
r_{H}^{2}=0.01$ (lower curves),$\;0.05$ (upper curves), for $n=0$, $1$, $2$
from left to right.}%
\label{fig6}%
\end{figure}

Now we turn to the entropy variation. One will find that it is not suitable to
integrate Eq. (\ref{dSrad}) directly because it can not be exactly computed
near zero frequency. For explicity, we plot the integrands in Fig. \ref{fig5},
where it is shown that the lowest mode $\left(  j=0\right)  $ of
integrands of entropy radiation tends to infinity as $\omega\rightarrow0$. It
can be seen clearly from Eq. (\ref{g}), where the term $\frac{\left\vert
A_{j}^{(s)}\right\vert ^{2}}{\exp(\omega/T_{H})-\left(  -1\right)  ^{2s}}$
diverges at zero frequency because the denominator is zero for $s=0$ and the
numerator $\left\vert A_{j}^{(s)}\right\vert ^{2}\neq0$ for $j=0$. Obviously,
we can not simply truncate the integration since we do not know whether the
truncated integration is finite and can be safely omitted. To tackle this
problem, we propose to evaluate the lowest mode near $\omega\rightarrow0$
analytically. Fortunately, this lowest mode has been found in \cite{SdS} as%
\[
\left\vert A_{0}^{(0)}\right\vert ^{2}=\frac{4\left(  r_{C}r_{H}\right)
^{n+2}}{\left(  r_{C}^{n+2}+r_{H}^{n+2}\right)  ^{2}}.
\]
Substituting this solution into Eq. (\ref{g}), one can find that the
integration is finite. Comparing the analytical integrand with the numerical
result in Fig. \ref{fig6}, one can find they accord very well for small $\omega$, so the
truncated integration can be safely performed. In Table \ref{tab1}, we list
the radiated entropy and lost entropy for different extra dimensions and the
cosmological constant. \begin{table}[th]%
\begin{tabular}
[c]{|c|c|c|c|c|c|c|c|c|c|c|c|c|}\hline
$\Lambda r_{H}^{2}$ & \multicolumn{4}{|c|}{$0$} & \multicolumn{4}{|c|}{$0.01$}
& \multicolumn{4}{|c|}{$0.05$}\\\hline
$n$ & $0$ & $1$ & $2$ & $3$ & $0$ & $1$ & $2$ & $3$ & $0$ & $1$ & $2$ &
$3$\\\hline
$dS_{rad}$ & 6.914$n_{0}$ & 30.19$n_{0}$ & 79.12$n_{0}$ & 160.6$n_{0}$ &
14.09$n_{0}$ & 36.87$n_{0}$ & 85.39$n_{0}$ & 166.0$n_{0}$ & 26.76$n_{0}$ &
48.59$n_{0}$ & 96.41$n_{0}$ & 177.3$n_{0}$\\\hline
$dS$ & 3.740$n_{0}$ & 16.73$n_{0}$ & 44.97$n_{0}$ & 92.30$n_{0}$ &
7.535$n_{0}$ & 19.98$n_{0}$ & 47.78$n_{0}$ & 94.98$n_{0}$ & 15.45$n_{0}$ &
25.86$n_{0}$ & 53.46$n_{0}$ & 99.12$n_{0}$\\\hline
$R$ & 1.849 & 1.805 & 1.759 & 1.740 & 1.870 & 1.845 & 1.787 & 1.748 & 1.732 &
1.879 & 1.803 & 1.788\\\hline
\end{tabular}
\caption{The entropy gained by radiation $dS_{rad}$, lost by black hole $dS$,
and their ratio, for scalar field on the brane in different cases with
$\Lambda r_{H}^{2}=0$, 0.01, 0.05 and $n=0$, 1, 2, 3, respectively. The unit
of entropy is 10$^{-3}r_{H}^{-1}$ and $n_{0}$ denotes the scalar degrees of
freedom. }%
\label{tab1}%
\end{table}

\subsection{fermion field ($s=1/2$)}

\begin{figure}[ptb]
\begin{center}
\includegraphics[
height=3in, width=6in ]{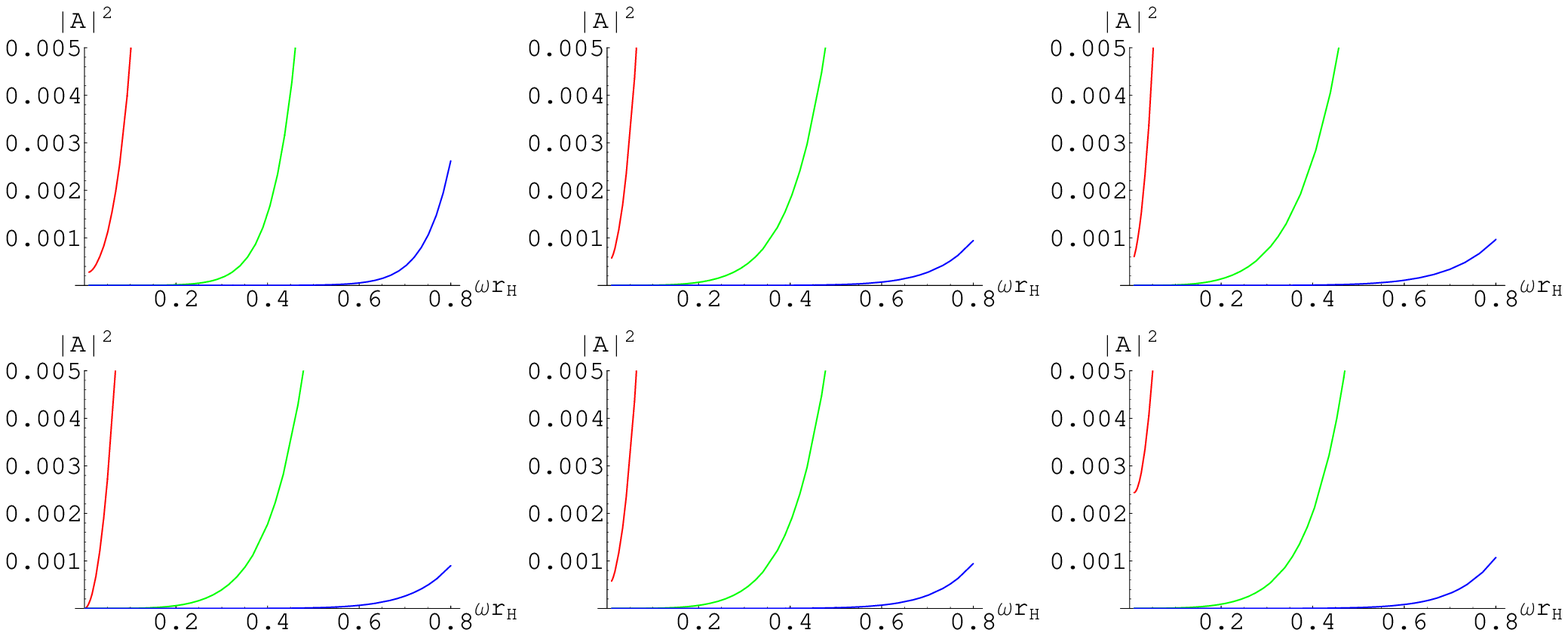}
\end{center}
\caption{Absorption probability $\left\vert A_{j}^{(\frac{1}{2})}\right\vert
^{2}$ of fermion field with $j$ $=\frac{1}{2}$ (red)$,\;\frac{3}{2}$
(green)$,\;\frac{5}{2}$\ (blue). For the top panel: $\Lambda r_{H}^{2}=0.01,$
and $n=0,\;1,\;2$ from left to right; for the bottom panel: $n=1$ and $\Lambda
r_{H}^{2}=0,\;0.01,\;0.05$ from left to right.}%
\label{fig7}%
\end{figure}

\begin{figure}[ptb]
\begin{center}
\includegraphics[
height=2in, width=6in ]{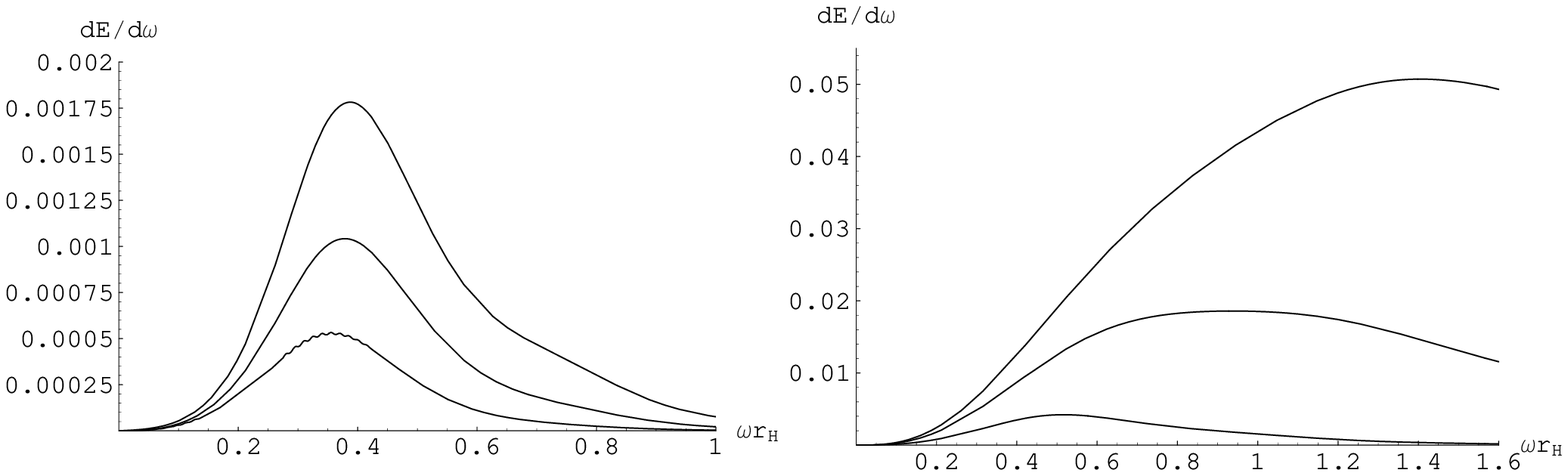}
\end{center}
\caption{The differential energy emission rate for fermion emission on the
brane. For the left panel: the dimensionality of spacetime is fixed at $n=0$,
while $\Lambda r_{H}^{2}$ takes the values $\{0,0.01,0.05\}$ from bottom to
top; for the right panel: the cosmological constant is fixed at $\Lambda
r_{H}^{2}$ $=0.01$, and $n$ takes the indicative values $\{1,3,5\}$ from
bottom to top.}%
\label{fig8}%
\end{figure}

To construct the complete solution for the emitted field with non-vanishing
spin, one should consider, in principle, both the upper and lower components
of the field. However, the determination of either is more than adequate to
compute the absorption coefficient. Considering the radial component of
conserved current for fermions $J^{u}=\sqrt{2}\sigma_{AB}^{u}\Psi^{A}\bar
{\Psi}^{B}$, Cveti\v{c} and Larsen\ \cite{Cvetic} have obtained the ingoing
flux%
\[
F=\left\vert P_{\frac{1}{2}}\right\vert ^{2}-\left\vert P_{-\frac{1}{2}%
}\right\vert ^{2}.
\]
We need to evaluate this formula at two horizons. Similar to the case for
Schwarzschild black hole, we find that, from the asymptotic solutions near
horizons (\ref{xh}) and (\ref{cx}), the upper and lower components of the
emitted field mainly carry the ingoing and outgoing waves respectively. Thus,
the ratio between the ingoing flux near two horizons may be directly written
as\ \cite{Cvetic,Kanti}%
\[
\left\vert A_{j}^{(\frac{1}{2})}\right\vert ^{2}=\frac{\left\vert P_{\frac
{1}{2}}\right\vert _{r\rightarrow r_{H}}^{2}}{\left\vert P_{\frac{1}{2}%
}\right\vert _{r\rightarrow r_{C}}^{2}}=\frac{\left\vert A_{1}\right\vert
_{r\rightarrow r_{H}}^{2}}{\left\vert B_{1}\right\vert _{r\rightarrow r_{C}%
}^{2}}.
\]
In Fig. \ref{fig7}, we plot the absorption probability of fermion field as function of
the energy parameters $\omega r_{H}$ with different $j$, $n$, and $\Lambda
r_{H}^{2}$. Similar to the scalar case, one can find that the lowest mode of
absorption probability is nonvanishing when $\omega\rightarrow0$ in the
presence of cosmological constant. Comparing Fig. \ref{fig3} and Fig.
\ref{fig7}, it can be noted that the nonvanishing absorption probability for
fermion field is about ten times smaller than that of the scalar field. The
corresponding energy spectrum is given in Fig. \ref{fig8}. It could be found that the energy emission rate is enhanced, when
the space-time dimension or the cosmological constant increases. The peak is
shifted towards higher energies when space-time dimension increases, but when
the cosmological constant increases, the peak is not shifted significantly.
These features are analogous to the behavior found in the asymptotically-flat
Schwarzschild space-time \cite{Kanti} and the scalar case shown in Fig.
\ref{fig4}. However, the energy emission rate always vanishes when
$\omega\rightarrow0$, which is different from the scalar case shown in Fig.
\ref{fig4}, but is the same as the one in asymptotically-flat Schwarzschild
space-time. Moreover, one can find that the integrands of entropy in Eqs.
(\ref{h}) and (\ref{g}) are not divergent when $\omega\rightarrow0$ because
the denominator $\exp(\omega/T_{H})-\left(  -1\right)  ^{2s}$ is nonvanishing
for $s=1/2$. Hence the integration can be directly calculated. In Table
\ref{tab2}, we list the radiated entropy and lost entropy for different extra
dimensions and cosmological constants. \begin{table}[th]%
\begin{tabular}
[c]{|c|c|c|c|c|c|c|c|c|c|c|c|c|}\hline
$\Lambda r_{H}^{2}$ & \multicolumn{4}{|c|}{$0$} & \multicolumn{4}{|c|}{$0.01$}
& \multicolumn{4}{|c|}{$0.05$}\\\hline
$n$ & $0$ & $1$ & $2$ & $3$ & $0$ & $1$ & $2$ & $3$ & $0$ & $1$ & $2$ &
$3$\\\hline
$dS_{rad}$ & 3.370$n_{\frac{1}{2}}$ & 24.41$n_{\frac{1}{2}}$ & 68.07$n_{\frac
{1}{2}}$ & 137.6$n_{\frac{1}{2}}$ & 6.020$n_{\frac{1}{2}}$ & 27.52$n_{\frac
{1}{2}}$ & 70.92$n_{\frac{1}{2}}$ & 140.2$n_{\frac{1}{2}}$ & 9.823$n_{\frac
{1}{2}}$ & 32.05$n_{\frac{1}{2}}$ & 75.73$n_{\frac{1}{2}}$ & 145.0$n_{\frac
{1}{2}}$\\\hline
$dS$ & 2.057$n_{\frac{1}{2}}$ & 14.57$n_{\frac{1}{2}}$ & 40.71$n_{\frac{1}{2}%
}$ & 82.72$n_{\frac{1}{2}}$ & 4.453$n_{\frac{1}{2}}$ & 17.23$n_{\frac{1}{2}}$
& 43.23$n_{\frac{1}{2}}$ & 85.46$n_{\frac{1}{2}}$ & 8.632$n_{\frac{1}{2}}$ &
21.34$n_{\frac{1}{2}}$ & 47.47$n_{\frac{1}{2}}$ & 89.72$n_{\frac{1}{2}}%
$\\\hline
$R$ & 1.638 & 1.676 & 1.672 & 1.664 & 1.352 & 1.597 & 1.641 & 1.641 & 1.138 &
1.502 & 1.595 & 1.616\\\hline
\end{tabular}
\caption{The two entropy variations and their ratio, for fermions on the brane
in different cases. $n_{1/2}$ denotes the fermionic degrees of freedom. }%
\label{tab2}%
\end{table}

\subsection{gauge boson fields ($s=1$)}

The calculation for gauge boson fields is more difficult, since the outgoing
mode in the upper component of the wave-function is very small, thus a tiny
error in the numerical solution can easily be mixed with the outgoing solution
and can consequently contaminate the ingoing one. As a result, the numerical
integration is not stable as it strongly depends on the boundary conditions. A
solution to overcome this problem has been proposed in Ref. \cite{Harris} and
consists in solving the equation of motion for a new unknown radial function
$P_{1}=yF(y)e^{-i\omega r_{\ast}}$ with $y=r/r_{H}$. In terms of these new
variables, the wave equation for gauge boson fields becomes%
\[
fy^{2}\frac{d^{2}F}{dy^{2}}+2y(F-i\omega r_{H}y)\frac{dF}{dy}-j(j+1)F=0,
\]
with the following boundary conditions:%
\[
F(1)=1,\;\left.  \frac{dF}{dy}\right\vert _{y=1}=\frac{ij(j+1)}{2\omega r_{H}%
}.
\]

The ingoing flux is derived from the trace of the energy-momentum tensor
$T^{uv}=2\sigma_{AA^{\prime}}^{u}\sigma_{BB^{\prime}}^{v}\Psi^{AB}\bar{\Psi
}^{A^{\prime}B^{\prime}}$ evaluated over a two dimensional sphere. It has been
got \cite{Cvetic}%
\[
F=\frac{1}{2r^{2}\omega}\left(  \left\vert P_{1}\right\vert ^{2}-\left\vert
P_{-1}\right\vert ^{2}\right)  .
\]
As for fermions, the absorption probability can be determined only by upper
component of the field. So we get%
\[
\left\vert A_{j}^{(1)}\right\vert ^{2}=\frac{\left\vert \frac{P_{1}}{r^{2}%
}\right\vert _{r\rightarrow r_{H}}^{2}}{\left\vert \frac{P_{1}}{r^{2}%
}\right\vert _{r\rightarrow r_{C}}^{2}}=\frac{1}{\left\vert F\right\vert
_{y\rightarrow r_{C}}^{2}}.
\]
The last equality has used the new radial wave function and boundary condition.

\begin{figure}[ptb]
\begin{center}
\includegraphics[
height=3in, width=6in ]{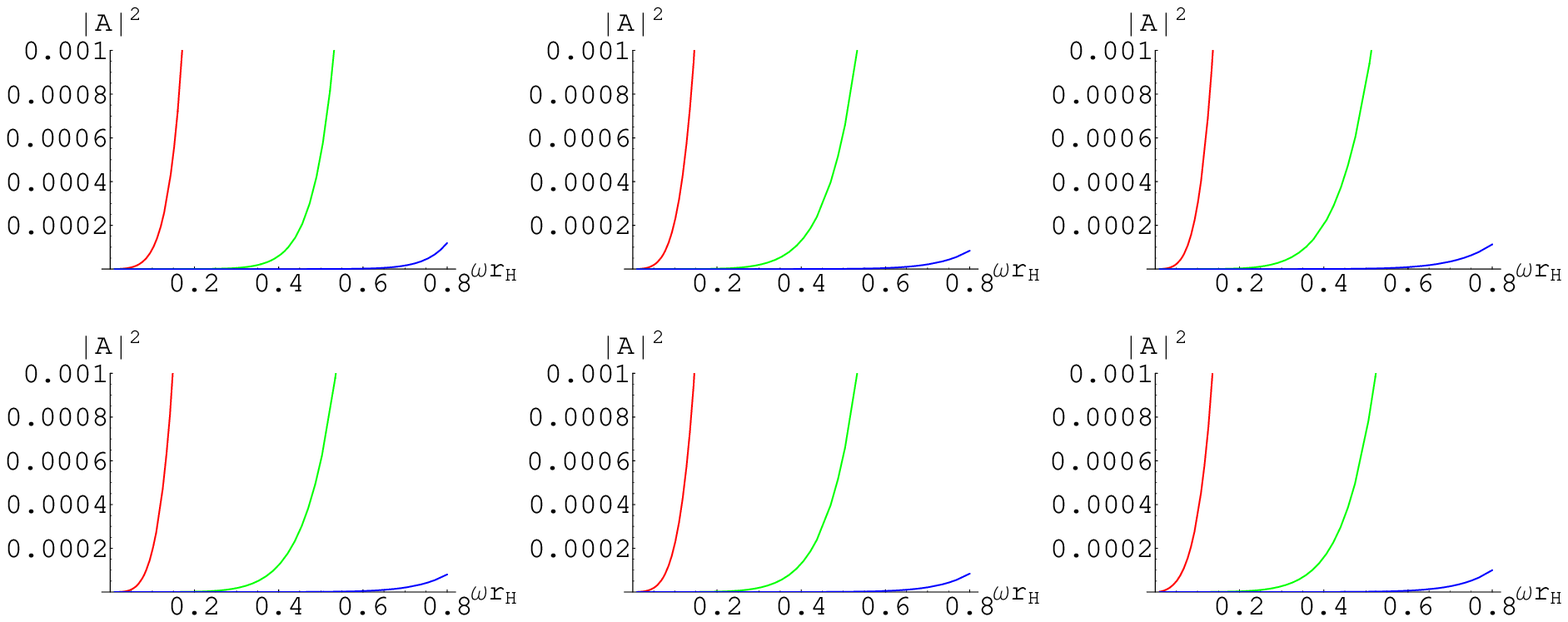}
\end{center}
\caption{Absorption probability $\left\vert A_{j}^{(1)}\right\vert ^{2}$ of
gauge boson field with $j$ $=1$ (red)$,\;2$ (green)$,\;3$\ (blue). For the top
panel: $\Lambda r_{H}^{2}=0.01,$ and $n=0,\;1,\;2$ from left to right; for the
bottom panel: $n=1$ and $\Lambda r_{H}^{2}=0,\;0.01,\;0.05$ from left to
right.}%
\label{fig9}%
\end{figure}

\begin{figure}[ptb]
\begin{center}
\includegraphics[
height=2in, width=6in ]{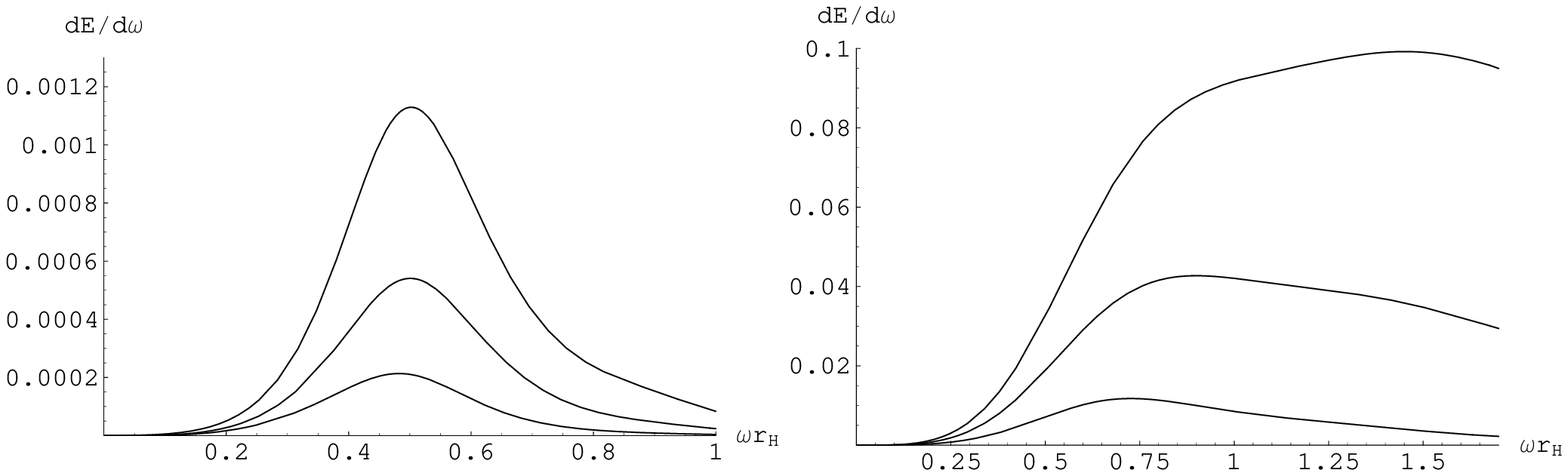}
\end{center}
\caption{The differential energy emission rate for gauge boson emission on the
brane. For the left panel: the dimensionality of space-time is fixed at $n=0$,
while $\Lambda r_{H}^{2}$ takes the values $\{0,0.01,0.05\}$ from bottom to
top; for the right panel: the cosmological constant is fixed at $\Lambda
r_{H}^{2}$ $=0.01$, and $n$ takes the values $\{2,4,6\}$ from bottom to top.}%
\label{fig10}%
\end{figure}

In Fig. \ref{fig9} and \ref{fig10}, we plot the absorption probability and
energy spectrum. Compared with the scalar particles and fermions, the absorption
probability of gauge bosons always vanishes when $\omega\rightarrow0$, which
ensures that the integrand of radiated entropy (\ref{g}) is not divergent when
$\omega\rightarrow0$ although the denominator $\exp(\omega/T_{H})-\left(
-1\right)  ^{2s}$ for $s=1$ vanishes as $\omega\rightarrow0$. In Table
\ref{tab3}, we list the radiated entropy and lost entropy. \begin{table}[th]%
\begin{tabular}
[c]{|c|c|c|c|c|c|c|c|c|c|c|c|c|}\hline
$\Lambda r_{H}^{2}$ & \multicolumn{4}{|c|}{$0$} & \multicolumn{4}{|c|}{$0.01$}
& \multicolumn{4}{|c|}{$0.05$}\\\hline
$n$ & $0$ & $1$ & $2$ & $3$ & $0$ & $1$ & $2$ & $3$ & $0$ & $1$ & $2$ &
$3$\\\hline
$dS_{rad}$ & 1.268$n_{1}$ & 17.87$n_{1}$ & 64.19$n_{1}$ & 147.5$n_{1}$ &
2.837$n_{1}$ & 20.82$n_{1}$ & 67.41$n_{1}$ & 150.6$n_{1}$ & 5.533$n_{1}$ &
25.30$n_{1}$ & 72.95$n_{1}$ & 156.5$n_{1}$\\\hline
$dS$ & 0.845$n_{1}$ & 11.47$n_{1}$ & 40.72$n_{1}$ & 93.23$n_{1}$ &
2.272$n_{1}$ & 13.98$n_{1}$ & 43.46$n_{1}$ & 95.89$n_{1}$ & 5.222$n_{1}$ &
17.98$n_{1}$ & 48.24$n_{1}$ & 101.0$n_{1}$\\\hline
$R$ & 1.500 & 1.557 & 1.576 & 1.582 & 1.249 & 1.489 & 1.551 & 1.570 & 1.060 &
1.407 & 1.512 & 1.549\\\hline
\end{tabular}
\caption{The two entropy variations and their ratio, for bosons on the brane
in different cases. $n_{1}$ denotes the bosonic degrees of freedom.}%
\label{tab3}%
\end{table}

\subsection{graviton}

\bigskip The absorption probability for the graviton modes can be obtained
similar to the scalar field case. We write it directly as%
\[
\left\vert A_{j}\right\vert ^{2}=1-\left\vert \frac{F_{2}}{F_{1}}\right\vert
^{2}=1-\left\vert \frac{W\left[  F_{-},\Phi\right]  }{W\left[  F_{+}%
,\Phi\right]  }\right\vert ^{2},
\]
where $F_{-}$ and $F_{+}$ denote the ingoing and outgoing waves in asymptotic
solution obtained in Eq. (\ref{cx1}).

\begin{figure}[ptb]
\begin{center}
\includegraphics[
height=3in, width=6in ]{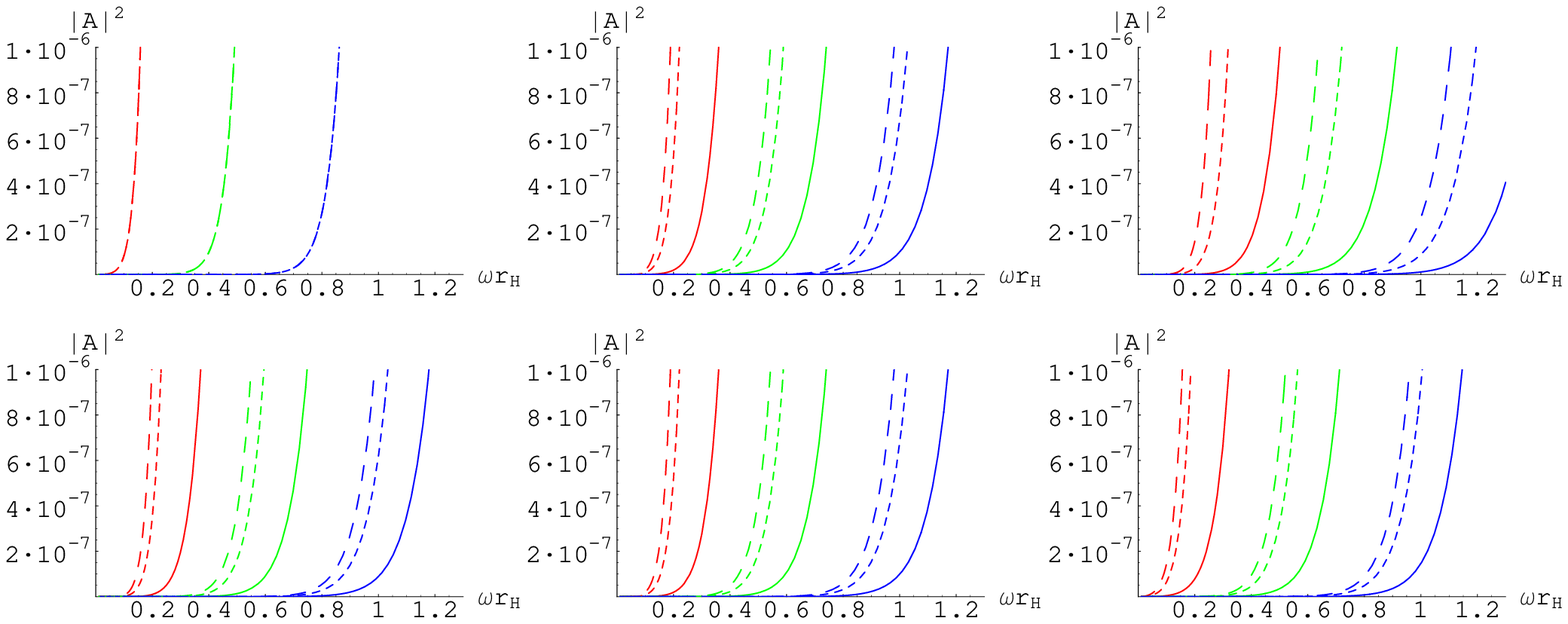}
\end{center}
\caption{Absorption probability $\left\vert A_{j}^{(2)}\right\vert ^{2}$ for
tensor (solid lines), vector (short-dashed lines) and scalar (long-dashed
lines) gravitational perturbations in the bulk with $j$ $=2$ (red)$,\;3$
(green)$,\;4$\ (blue). For the top panel: $\Lambda r_{H}^{2}=0.01,$ and
$n=0,\;1,\;2$ from left to right; for the bottom panel: $n=1$ and $\Lambda
r_{H}^{2}=0,\;0.01,\;0.05$ from left to right.}%
\label{fig11}%
\end{figure}

\begin{figure}[ptb]
\begin{center}
\includegraphics[
height=3in, width=6in ]{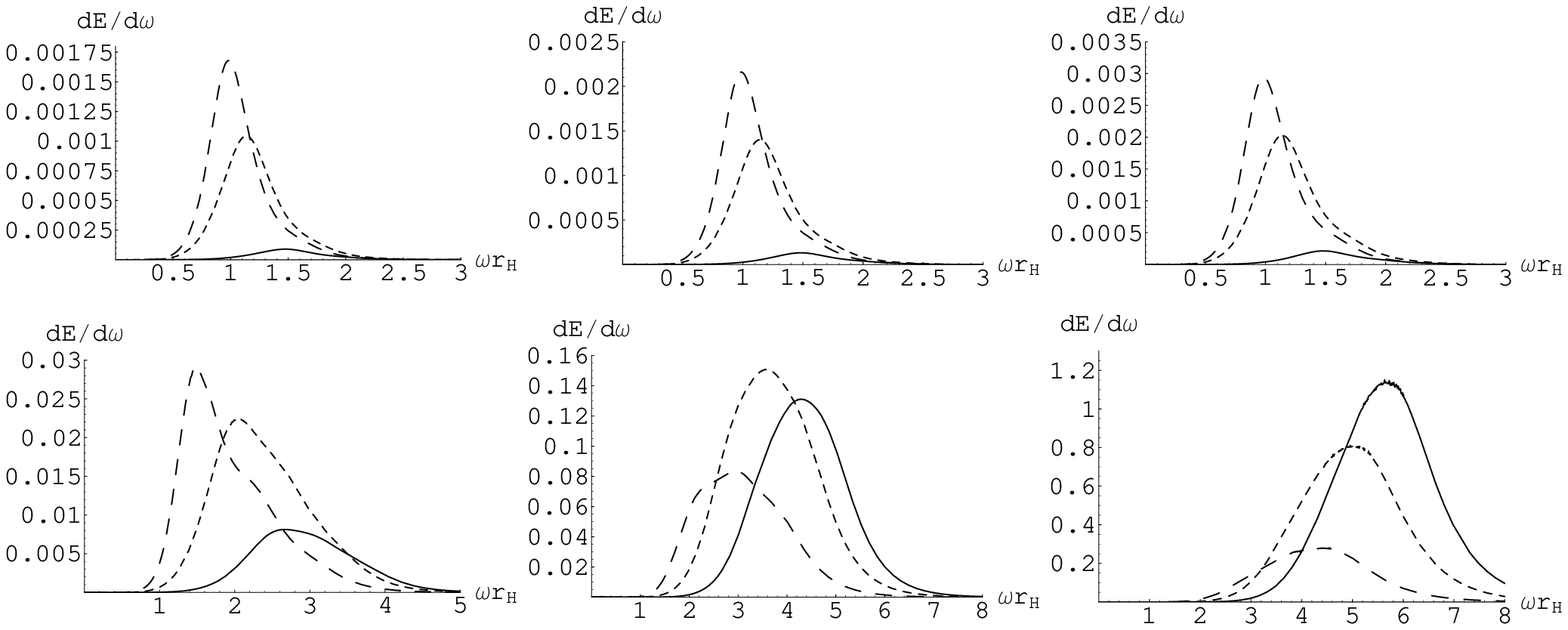}
\end{center}
\caption{The differential energy emission rate for tensor (solid lines),
vector (short-dashed lines) and scalar (long-dashed lines) gravitational
perturbations in the bulk. For the top panel: $n=1$, while $\Lambda r_{H}^{2}
$ takes the values $\{0,0.01,0.05\}$ from left to right; for the bottom panel:
$\Lambda r_{H}^{2}$ $=0.01$, and $n$ takes the values $\{3,5,7\}$ from left to
right.}%
\label{fig12}%
\end{figure}

In Fig. \ref{fig11}, we plot the absorption probability of the three perturbations. One
can find that the probability for the scalar mode graviton is always the
biggest one, followed by the vector mode and tensor mode. It is known that, in
the 4d limit, the vector mode corresponds to the gravitational axial
perturbation \cite{Vishveshwara}, and the scalar mode corresponds to the
gravitational polar perturbation \cite{Zerilli}. There is no counterpart of
the tensor mode in four dimension. An interesting feature is that the
absorption probability for the vector mode and scalar mode of graviton by a
SSdS black hole are exactly the same in four dimensions, which is similar in
the case of Schwarzschild black hole \cite{Chandrasekhar,Park}. The
corresponding energy spectrum is plotted in Fig. \ref{fig12}. We find that the
energy emission rate is enhanced for all modes when the space-time dimensions
and cosmological constant increase. The peak is shifted towards higher
frequency clearly for all modes when space-time dimensions increase. Moreover,
we find that the energy emission rate of tensor mode is enhanced more
obviously than vector and scalar modes when space-time dimensions increase.
Similar to the case of gauge bosons, the integrand of radiated
entropy is not divergent when\textbf{\ }$\omega\rightarrow0$\textbf{.} In
Table \ref{tab4}, \begin{table}[th]%
\begin{tabular}
[c]{|c|c|c|c|c|c|c|c|c|c|c|c|}\hline
$n$ & \multicolumn{2}{|c|}{0} & \multicolumn{3}{|c|}{1} &
\multicolumn{3}{|c|}{2} & \multicolumn{3}{|c|}{3}\\\hline
mode & scalar & vector & scalar & vector & tensor & scalar & vector & tensor &
scalar & vector & tensor\\\hline\hline
$\Lambda r_{H}^{2}$ & \multicolumn{11}{|c|}{0}\\\hline
$dS_{rad}$ & 0.130$n_{2s}$ & 0.130$n_{2v}$ & 7.154$n_{2s}$ & 5.403$n_{2v}$ &
0.538$n_{2t}$ & 45.51$n_{2s}$ & 35.73$n_{2v}$ & 8.041$n_{2t}$ & 135.2$n_{2s}$
& 136.3$n_{2v}$ & 54.21$n_{2t}$\\\hline
$dS$ & 0.097$n_{2s}$ & 0.097$n_{2v}$ & 5.334$n_{2s}$ & 3.969$n_{2v}$ &
0.404$n_{2t}$ & 33.70$n_{2s}$ & 26.43$n_{2v}$ & 6.138$n_{2t}$ & 101.3$n_{2s}$
& 102.5$n_{2v}$ & 42.20$n_{2t}$\\\hline
$R$ & 1.347 & 1.347 & 1.341 & 1.361 & 1.331 & 1.350 & 1.352 & 1.310 & 1.334 &
1.330 & 1.285\\\hline
\end{tabular}
\begin{tabular}
[c]{|c|c|c|c|c|c|c|c|c|c|c|c|}\hline
$\Lambda r_{H}^{2}$ & \multicolumn{11}{|c|}{0.01}\\\hline
$dS_{rad}$ & 0.447$n_{2s}$ & 0.447$n_{2v}$ & 9.062$n_{2s}$ & 7.097$n_{2v}$ &
0.768$n_{2t}$ & 48.61$n_{2s}$ & 39.43$n_{2v}$ & 9.087$n_{2t}$ & 140.5$n_{2s}$
& 143.0$n_{2v}$ & 57.63$n_{2t}$\\\hline
$dS$ & 0.397$n_{2s}$ & 0.397$n_{2v}$ & 7.050$n_{2s}$ & 5.451$n_{2v}$ &
0.604$n_{2t}$ & 37.04$n_{2s}$ & 29.64$n_{2v}$ & 7.053$n_{2t}$ & 106.0$n_{2s}$
& 108.4$n_{2v}$ & 45.18$n_{2t}$\\\hline
$R$ & 1.125 & 1.125 & 1.285 & 1.302 & 1.271 & 1.312 & 1.330 & 1.288 & 1.326 &
1.320 & 1.276\\\hline
\end{tabular}
\begin{tabular}
[c]{|c|c|c|c|c|c|c|c|c|c|c|c|}\hline
$\Lambda r_{H}^{2}$ & \multicolumn{11}{|c|}{0.05}\\\hline
$dS_{rad}$ & 1.229$n_{2s}$ & 1.229$n_{2v}$ & 12.23$n_{2s}$ & 10.11$n_{2v}$ &
1.222$n_{2t}$ & 55.36$n_{2s}$ & 46.64$n_{2v}$ & 11.33$n_{2t}$ & 151.1$n_{2s}$
& 157.1$n_{2v}$ & 64.77$n_{2t}$\\\hline
$dS$ & 1.283$n_{2s}$ & 1.283$n_{2v}$ & 10.05$n_{2s}$ & 8.223$n_{2v}$ &
1.022$n_{2t}$ & 43.23$n_{2s}$ & 36.00$n_{2v}$ & 9.037$n_{2t}$ & 115.5$n_{2s}$
& 120.7$n_{2v}$ & 51.51$n_{2t}$\\\hline
$R$ & 0.957 & 0.957 & 1.217 & 1.229 & 1.195 & 1.281 & 1.296 & 1.253 & 1.308 &
1.302 & 1.257\\\hline
\end{tabular}
\caption{The two entropy variations and their ratio, for gravitons in the bulk
in different cases. $n_{2s}$, $n_{2v}$, and $n_{2t}$ denote the gravitational
degrees of freedom for scalar $\left(  n_{2s}\right)  $, vector $\left(
n_{2v}\right)  $, and tensor $\left(  n_{2t}\right)  $ perturbations,
respectively.}%
\label{tab4}%
\end{table}we list the radiated entropy and lost entropy for different
perturbations, cosmological constant, and the numbers of extra dimensions. One
may have noticed a strange ratio $R=0.957<1$ for four-dimensional gravitons
with large cosmological constant $\Lambda r_{H}^{2}=0.05$. This is
unreasonable since it destroys the generalized second law of thermodynamics.
In fact, we find that, if the cosmological constant is big and the space-time
dimensions are small enough, i.e. the black hole is close to the Nariai black
hole, the ratio $R$ may be smaller than unit for all types of particles. For
example, the ratio $R$ is $0.931$ for bosons with $n=0$ and $\Lambda r_{H}%
^{2}=0.1$. Hence, there should be something wrong near the Nariai limit. In
the recent work \cite{Myung2} on the thermal stability of Nariai black hole,
it was pointed out that black hole thermodynamics favors the standard Hawking
temperature rather than the Bousso-Hawking temperature, because the later is
inappropriate to describe either the Hawking-Page phase transition or the
evaporation process. We evaluate the ratio $R$ using the standard temperature
and find that the ratio is always bigger than 1. For examples, the ratio for
gravitons with $n=0$ and $\Lambda r_{H}^{2}=0.05$ now is $R=1.376$, and the
ratio for bosons with $n=0$ and $\Lambda r_{H}^{2}=0.1$ now is $R=1.407$. We
show the comparison between these two temperatures with some indicative cases
near the Nariai limit in Table \ref{tab5}. These results favor the argument
given in \cite{Myung2} but obtained from a different aspect of thermodynamics.
\begin{table}[th]%
\begin{tabular}
[c]{|c|c|c|c|c|c|c|c|c|c|c|c|c|c|}\hline
\multicolumn{2}{|c}{$n$ and $s$} & \multicolumn{3}{|c|}{n=0\ s=2 (vector
mode)\ $\left(  \lambda_{0}=1\right)  $} & \multicolumn{3}{|c|}{n=1
s=1\ $\left(  \lambda_{1}=3\right)  $} & \multicolumn{3}{|c|}{n=2 s=$\frac
{1}{2}$\ $\left(  \lambda_{2}=6\right)  $} & \multicolumn{3}{|c|}{n=3
s=0\ $\left(  \lambda_{3}=10\right)  $}\\\hline
\multicolumn{2}{|c|}{$\Lambda r_{H}^{2}$} & 0.1$\lambda_{0}$ & 0.5$\lambda
_{0}$ & 0.9$\lambda_{0}$ & 0.1$\lambda_{1}$ & 0.5$\lambda_{1}$ &
0.9$\lambda_{1}$ & 0.1$\lambda_{2}$ & 0.5$\lambda_{2}$ & 0.9$\lambda_{2}$ &
0.1$\lambda_{3}$ & 0.5$\lambda_{3}$ & 0.9$\lambda_{3}$\\\hline\hline
& $dS_{rad}$ & 2.456$n_{2v}$ & 49.51$n_{2v}$ & 2215$n_{2v}$ & 43.34$n_{1}$ &
187.2$n_{1}$ & 3547$n_{1}$ & 92.72$n_{\frac{1}{2}}$ & 294.7$n_{\frac{1}{2}}$ &
4111$n_{\frac{1}{2}}$ & 293.7$n_{0}$ & 710.0$n_{0}$ & 6208$n_{0}%
$\\\cline{2-14}%
$T_{H}$ & $dS$ & 2.908$n_{2v}$ & 130.9$n_{2v}$ & 29077$n_{2v}$ & 36.80$n_{1}$
& 335.8$n_{1}$ & 36453$n_{1}$ & 63.79$n_{\frac{1}{2}}$ & 454.9$n_{\frac{1}{2}%
}$ & 35043$n_{\frac{1}{2}}$ & 181.4$n_{0}$ & 852.0$n_{0}$ & 45156$n_{0}%
$\\\cline{2-14}
& $R$ & 0.844 & 0.378 & 0.0762 & 1.178 & 0.558 & 0.0973 & 1.453 & 0.648 &
0.117 & 1.619 & 0.833 & 0.138\\\hline\hline
& $dS_{rad}$ & 0.114$n_{2v}$ & 0.0486$n_{2v}$ & 0.00589$n_{2v}$ & 16.17$n_{1}
$ & 7.972$n_{1}$ & 1.182$n_{1}$ & 64.29$n_{\frac{1}{2}}$ & 31.69$n_{\frac
{1}{2}}$ & 5.460$n_{\frac{1}{2}}$ & 194.9$n_{0}$ & 137.8$n_{0}$ & 26.61$n_{0}%
$\\\cline{2-14}%
$T_{0}$ & $dS$ & 0.0809$n_{2v}$ & 0.0295$n_{2v}$ & 0.00304$n_{2v}$ &
9.916$n_{1}$ & 4.387$n_{1}$ & 0.604$n_{1}$ & 37.84$n_{\frac{1}{2}}$ &
15.68$n_{\frac{1}{2}}$ & 2.345$n_{\frac{1}{2}}$ & 95.90$n_{0}$ & 68.03$n_{0}$
& 13.35$n_{0}$\\\cline{2-14}
& $R$ & 1.405 & 1.645 & 1.935 & 1.630 & 1.817 & 1.959 & 1.699 & 2.021 &
2.328 & 2.032 & 2.025 & 1.993\\\hline
\end{tabular}
\caption{The two entropy variations and their ratio for the Bousso-Hawking
temperature $T_{H}$ and the standard Hawking temperature $T_{0}$. $\lambda
_{n}$ denotes the cosmological constant $\Lambda r_{H}^{2}$ when the Nariai
limit $r_{H}=r_{C}$ arises for different space-time dimensions $n$.}%
\label{tab5}%
\end{table}

\section{Conclusion and discussion}

In this paper, we have studied the radiation of ($4+n$)-dimensional braneworld
black hole imbedded in the space-time with a positive cosmological constant.
We calculate the greybody factor and energy spectrum of Hawking radiation for
all types of particles, including scalars, fermions, gauge bosons, and the
gravitons with three modes. Until now, the exact spectrums have not been
obtained except for scalars and gravitons at low and asymptotic frequency.
Since scalar particles are rather elusive to be detected, the present analysis
on the spectrum for other particles is important. We first studied the
greybody factor and find that for fermions, the factor for the lowest mode is
nonvanishing in the low-energy limit, similar to the scalar field \cite{SdS},
while different from the cases of bosons and gravitons. The energy emission is
found to increase significantly for all particles with the increasing
cosmological constant and extra dimensions. And the peak of the spectrum
shifts towards high energy when space-time dimensions increase, however, it is
not shifted significantly with the increasing cosmological constant. These
results show that it is possible to detect two parameters from the energy
spectrum of braneworld SSdS black hole. However, compared with the
nonvanishing energy emission rate of scalar fields in the low-energy limit,
the energy emission rates for other particles vanish as $w\rightarrow0$.
Therefore, it is difficult to detect the emission of ultra-soft quanta by SSdS
black hole which is expected in Ref. \cite{SdS}.

Based on the knowledge on the greybody factor, we have calculated the entropy
lost by the black hole and gained by the radiation. We show that the standard
Hawking temperature is appropriate to describe the entropy radiation near the
Nariai limit, rather than the Bousso-Hawking temperature which is disfavored
by the generalized second law of thermodynamics. We find that the ratios of
two entropies are near unit for all fields. It should be emphasized that this
result can not be foreseen easily, especially for the scalar field, where the
integrand of radiated entropy is divergent in the low-energy limit while the
integrand of the lost entropy is finite. Our results strongly favor
Bekenstein's conjecture, even if the extra dimensions and cosmological
constant exist. It further assures us the deep relationship between the
gravitational entropy and the statistical entropy and is useful to fully
understand the entropy of braneworld SSdS black hole.

\begin{acknowledgments}
Shao-Feng Wu wishes to thank Prof. Bin Wang for the illuminating discussions.
This work was supported by NSFC under Grant Nos. 10575068 and 10604024, the
Shanghai Research Foundation No. 07dz22020, the CAS Knowledge Innovation
Project Nos. KJcx.syw.N2, the Shanghai Education Development Foundation, and
the Innovation Foundation of Shanghai University.
\end{acknowledgments}

\end{document}